# Facial Recognition Technology: An analysis with scope in India

Dr.S.B.Thorat
Director,
Institute of Technology and Mgmt
Nanded, Dist. - Nanded. (MS),
India
suryakant_thorat@yahoo.com

S.K.Nayak
Head, Dept. of Computer Science
Bahirji Smarak Mahavidyalaya,
Basmathnagar, Dist. - Hingoli. (MS),
India
sunilnayak1234@yahoo.com

Miss.Jyoti P Dandale
Lecturer
Institute of Technology and Mgmt
Nanded, Dist. - Nanded. (MS),
India
j.dandale@gmail.com

*Abstract*— **A facial recognition system is a computer application for automatically identifying or verifying a person from a digital image or a video frame from a video source. One of the way is to do this is by comparing selected facial features from the image and a facial database.It is typically used in security systems and can be compared to other biometrics such as fingerprint or eye iris recognition systems.**

**In this paper we focus on 3-D facial recognition system and biometric facial recognision system. We do critics on facial recognision system giving effectiveness and weaknesses. This paper also introduces scope of recognision system in India.**

## I. INTRODUCTION

Pioneers of Automated Facial Recognition include: Woody Bledsoe, Helen Chan Wolf, and Charles Bisson.

During 1964 and 1965, Bledsoe, along with Helen Chan and Charles Bisson, worked on using the computer to recognize human faces (Bledsoe 1966a, 1966b; Bledsoe and Chan 1965). He was proud of this work, but because the funding was provided by an unnamed intelligence agency that did not allow much publicity, so little of the work was published. Given a large database of images (in effect, a book of mug shots) and a photograph, the problem was to select from the database a small set of records such that one of the image records matched the photograph. The success of the method could be measured in terms of the ratio of the answer list to the number of records in the database. Bledsoe (1966a) described the following difficulties:

"This recognition problem is made difficult by the great variability in head rotation and tilt, lighting intensity and angle, facial expression, aging, etc. Some other attempts at facial recognition by machine have allowed for little or no variability in these quantities. Yet the method of correlation (or pattern matching) of unprocessed optical data, which is often used by some researchers, is certain to fail in cases where the variability is great. In particular, the correlation is very low between two pictures of the same person with two different head rotations".

This project was labeled man-machine because the human extracted the coordinates of a set of features from the photographs, which were then used by the computer for recognition. Using a graphics tablet (GRAFACON or RAND TABLET), the operator would extract the coordinates of features such as the center of pupils, the inside corner of eyes, the outside corner of eyes, point of widows peak, and so on. From these coordinates, a list of 20 distances, such as width of mouth and width of eyes, pupil to pupil, were computed. These operators could process about 40 pictures an hour. When building the database, the name of the person in the photograph was associated with the list of computed distances and stored in the computer. In the recognition phase, the set of distances was compared with the corresponding distance for each photograph, yielding a distance between the photograph and the database record. The closest records are returned.

This brief description is an oversimplification that fails in general because it is unlikely that any two pictures would match in head rotation, lean, tilt, and scale (distance from the camera). Thus, each set of distances is normalized to represent the face in a frontal orientation. To accomplish this normalization, the program first tries to determine the tilt, the lean, and the rotation. Then using these angles, the computer undoes the effect of these transformations on the computed distances. To compute these angles, the computer must know the three-dimensional geometry of the head. Because the actual heads were unavailable Bledsoe (1964) used a standard head derived from measurements on seven heads.

After Bledsoe left PRI in 1966, this work was continued at the Stanford Research Institute, primarily by Peter Hart. In experiments performed on a database of over 2000 photographs, the computer consistently outperformed humans when presented with the same recognition tasks (Bledsoe 1968). Peter Hart (1996) enthusiastically recalled the project with the exclamation, "It really worked!"

By about 1997, the system developed by Christoph von der Malsburg and graduate students of the University of Bochum in Germany and the University of Southern California in the United States outperformed most systems with those of Massachusetts Institute of Technology and the University of Maryland rated next. The Bochum system was developed through funding by the United States Army Research Laboratory. The software was sold as ZN-Face and used by customers such as Deutsche Bank and operators of airports and other busy locations. The software was "robust enough to make identifications from less-than-perfect face views. It can also often see through such impediments to identification as mustaches, beards, changed hair styles and glasses—even sunglasses".

In about January 2007, image searches were "based on the text surrounding a photo," for example, if text nearby mentions the image content. Polar Rose technology can guess from a photograph, in about 1.5 seconds, why any individual may look like in three dimensions, and thought they "will ask users to input the names of people they recognize in photos online" to help build a database.

## II. FACIAL TECHNOLOGY AT A GLANCE

Identix®, a company based in Minnesota, is one of many developers of facial recognition technology. Its software, FaceIt®, can pick someone's face out of a crowd, extract the face from the rest of the scene and compare it to a database of stored images. In order for this software to work, it has to know how to differentiate between a basic face and the rest of the background. Facial recognition software is based on the ability to recognize a face and then measure the various features of the face.

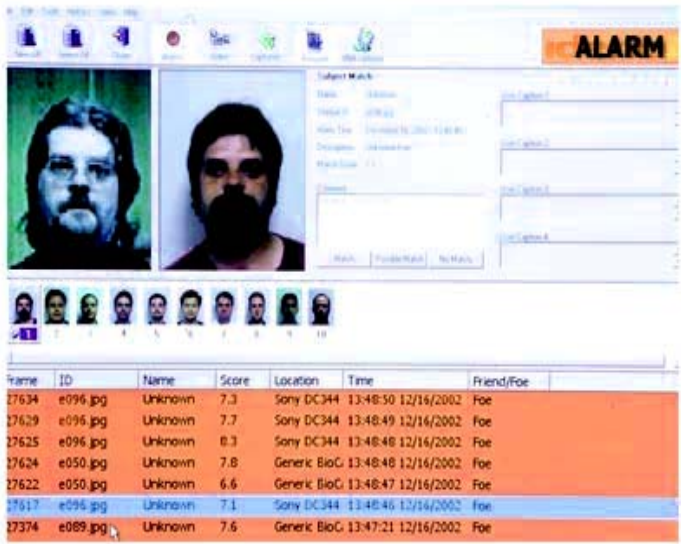

Figure 1.   Face IT software compares the face print with other images in the database. (Photo  Identix Inc.)

Every face has numerous, distinguishable landmarks, the different peaks and valleys that make up facial features. FaceIt defines these landmarks as nodal points. Each human face has approximately 80 nodal points. Some of these measured by the software are:

- Distance between the eyes
- Width of the nose
- Depth of the eye sockets
- The shape of the cheekbones
- The length of the jaw line

These nodal points are measured creating a numerical code, called a face print, representing the face in the database.

In the past, facial recognition software has relied on a 2D image to compare or identify another 2D image from the database. To be effective and accurate, the image captured needed to be of a face that was looking almost directly at the camera, with little variance of light or facial expression from the image in the database. This created quite a problem.

In most instances the images were not taken in a controlled environment. Even the smallest changes in light or orientation could reduce the effectiveness of the system, so they couldn't be matched to any face in the database, leading to a high rate of failure. In the next section, we will look at ways to correct the problem.

### A.   3D Facial Recognition

A newly-emerging trend in facial recognition software uses a 3D model, which claims to provide more accuracy. Capturing a real-time 3-D image of a person's facial surface, 3D facial recognition uses distinctive features of the face -- where rigid tissue and bone is most apparent, such as the curves of the eye socket, nose and chin -- to identify the subject. These areas are all unique and don't change over time.

Using depth and an axis of measurement that is not affected by lighting, 3D facial recognition can even be used in darkness and has the ability to recognize a subject at different view angles with the potential to recognize up to 90 degrees (a face in profile).

Using the 3D software, the system goes through a series of steps to verify the identity of an individual.

*a)   Detection:-* Acquiring an image can be accomplished by digitally scanning an existing photograph (2D) or by using a video image to acquire a live picture of a subject (3D).

*b)   Alignment:-* Once it detects a face, the system determines the head's position, size and pose. As stated earlier, the subject has the potential to be recognized up to 90 degrees. While with 2-D the head must be turned at least 35 degrees toward the camera.

*c)   Measurement:-* The system then measures the curves of the face on a sub-millimeter (or microwave) scale and creates a template.

*d)   Representation:-* The system translates the template into a unique code. This coding gives each template a set of numbers to represent the features on a subject's face.

*e)   Matching :-* If the image is 3D and the database contains 3D images, then matching will take place without any changes being made to the image. However, there is a challenge currently facing databases that are still in 2D images. 3D provides a live, moving variable subject being compared to a flat, stable image. New technology is addressing this challenge. When a 3D image is taken, different points (usually three) are identified. For example, the outside of the eye, the inside of the eye and the tip of the nose will be pulled out and measured.

Once those measurements are in place, an algorithm (a step-by-step procedure) will be applied to the image to convert it to a 2D image. After conversion, the software will then compare the image with the 2D images in the database to find a potential match.

*f) Verification or Identification* :- In verification, an image is matched to only one image in the database (1:1). For example, an image taken of a subject may be matched to an image in the Department of Motor Vehicles database to verify the subject is who he says he is. If identification is the goal, then the image is compared to all images in the database resulting in a score for each potential match (1:N). In this instance, you may take an image and compare it to a database of mug shots to identify who the subject is. Next, we'll look at how skin biometrics can help verify matches.

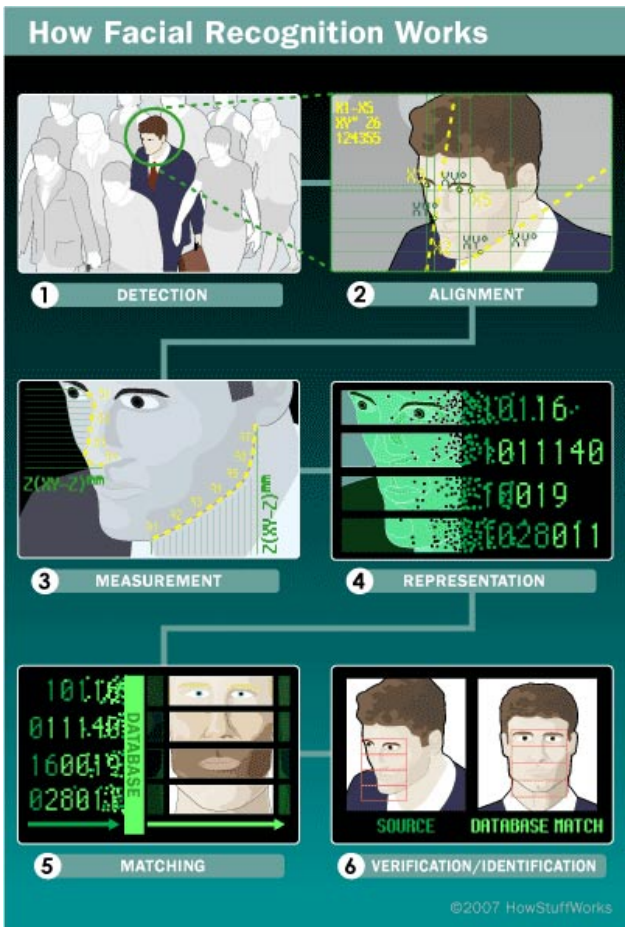

Figure 2. Working of facial recognisation

## B. Biometric Facial Recognition

The image may not always be verified or identified in facial recognition alone. Identix® has created a new product to help with precision. The development of FaceIt®Argus uses skin biometrics, the uniqueness of skin texture, to yield even more accurate results.

The process, called Surface Texture Analysis, works much the same way facial recognition does. A picture is taken of a patch of skin, called a skin print. That patch is then broken up into smaller blocks. Using algorithms to turn the patch into a mathematical, measurable space, the system will then distinguish any lines, pores and the actual skin texture. It can identify differences between identical twins, which is not yet possible using facial recognition software alone. According to Identix, by combining facial recognition with surface texture analysis, accurate identification can increase by 20 to 25 percent.

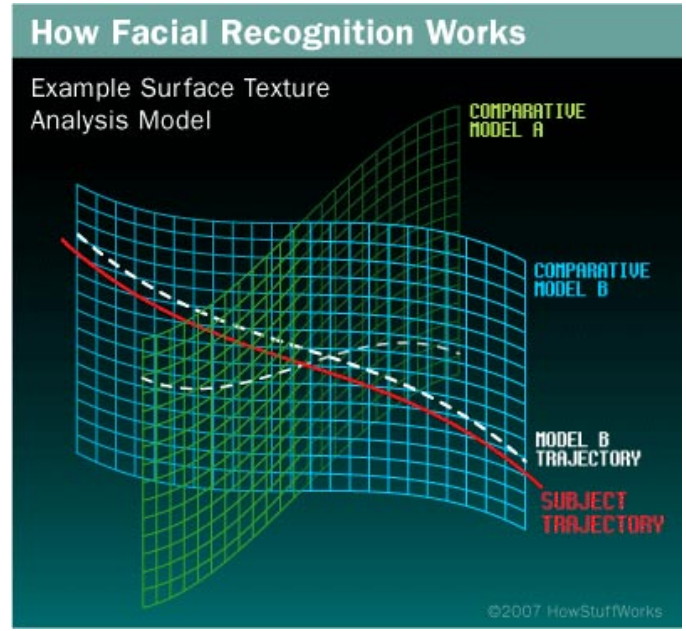

Figure 3. Working of facial recognisation (Surface texture analysis model)

FaceIt currently uses three different templates to confirm or identify the subject: vector, local feature analysis and surface texture analysis.

- The vector template is very small and is used for rapid searching over the entire database primarily for one-to-many searching.

- The Local Feature Analysis (LFA) template performs a secondary search of ordered matches following the vector template.

- The Surface Texture Analysis (STA) is the largest of the three. It performs a final pass after the LFA template search, relying on the skin features in the image, which contains the most detailed information.

By combining all three templates, FaceIt® has an advantage over other systems. It is relatively insensitive to changes in expression, including blinking, frowning or smiling and has the ability to compensate for mustache or beard growth and the appearance of eyeglasses. The system is also uniform with respect to race and gender.

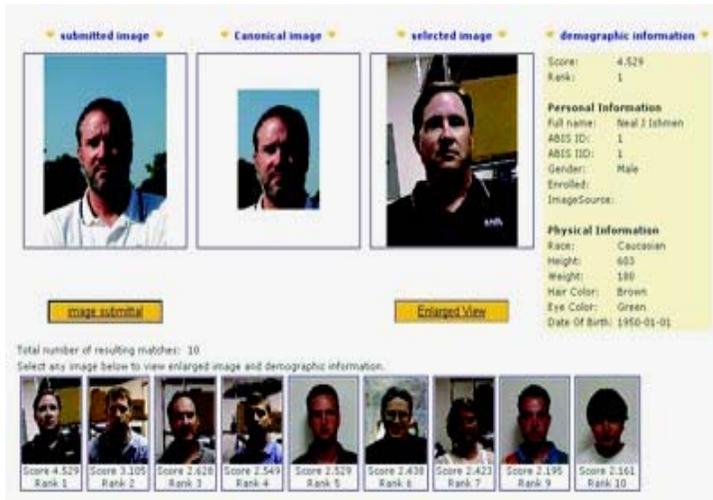

Figure 4.  Poor lighting can make it more difficult for facial recognition software to verify or identify someone.

However, it is not a perfect system. There are some factors that could get in the way of recognition, including:

- Significant glare on eyeglasses or wearing sunglasses.

- Long hair obscuring the central part of the face.

- Poor lighting that would cause the face to be over- or under-exposed.

- Lack of resolution (image was taken too far away).

Identix isn't the only company with facial recognition systems available. While most work the same way FaceIt does, there are some variations. For example, a company called Animetrix, Inc. has a product called FACEngine ID® SetLight that can correct lighting conditions that cannot normally be used, reducing the risk of false matches. Sensible Vision, Inc. has a product that can secure a computer using facial recognition. The computer will only power on and stay accessible as long as the correct user is in front of the screen. Once the user moves out of the line of sight, the computer is automatically secured from other users.

Due to these strides in technology, facial and skin recognition systems are more widely used than just a few years ago. In the next section, we'll look at where and how they are being used and what's in store for the future.

Among the different biometric techniques facial recognition may not be the most reliable and efficient but its great advantage is that it does not require aid from the test subject. Properly designed systems installed in airports, multiplexes, and other public places can identify individuals among the crowd. Other biometrics like fingerprints, iris, and speech recognition cannot perform this kind of mass scanning. However, questions have been raised on the effectiveness of facial recognition software in cases of railway and airport security.

## III.  CRITICISM

### A.  Weaknesses

Face recognition is not perfect and struggles to perform under certain conditions. Ralph Gross, a researcher at the Carnegie Mellon Robotics Institute, describes one obstacle related to the viewing angle of the face: "Face recognition has been getting pretty good at full frontal faces and 20 degrees off, but as soon as you go towards profile, there've been problems."

Other conditions where face recognition does not work well include poor lighting, sunglasses, long hair, or other objects partially covering the subject's face, and low resolution images.

Another serious disadvantage is that many systems are less effective if facial expressions vary. Even a big smile can render in the system less effective. For instance: Canada now allows only neutral facial expressions in passport photos.

### B.  Effectiveness

Critics of the technology complain that the London Borough of Newham scheme has, as of 2004, never recognized a single criminal, despite several criminals in the system's database living in the Borough and the system having been running for several years. "Not once, as far as the police know, has Newham's automatic facial recognition system spotted a live target." This information seems to conflict with claims that the system was credited with a 34% reduction in crime - which better explains why the system was then rolled out to Birmingham also.

An experiment by the local police department in Tampa, Florida, had similarly disappointing results.

"Camera technology designed to spot potential terrorists by their facial characteristics at airports failed its first major test at Boston's Logan Airport".

Safehouse International Limited, an Australian company, patented software including iMotion and iCount systems. The company claimed this system were able to track moving people and calculate the number of people in a crowd. After 9/11, the software was considered "commercially attractive" by the US administration. It was later revealed by David Mapley, a US shareholder of Safehouse International Limited) that the software actually never worked.

### C.  Privacy concerns

Despite the potential benefits of this technology, many citizens are concerned that their privacy will be invaded. Some fear that it could lead to a "total surveillance society," with the government and other authorities having the ability to know where you are, and what you are doing, at all times. This is not to be an underestimated concept as history has shown that states have typically abused such access before.

### D.  Recent improvements

In 2006, the performances of the latest face recognition algorithms were evaluated in the Face Recognition Grand Challenge (FRGC). High-resolution face images, 3-D face scans, and iris images were used in the tests. The results indicated that the new algorithms are 10 times more accurate than the face recognition algorithms of 2002 and 100 times more accurate than those of 1995. Some of the algorithms were able to outperform human participants in recognizing faces and could uniquely identify identical twins.

Low-resolution images of faces can be enhanced using face hallucination. Further improvements in high resolution, megapixel cameras in the last few years have helped to resolve the issue of insufficient resolution.

### E. Future development

**A possible future application for facial recognition systems lies in retailing.** A retail store (for example, a grocery store) may have cash registers equipped with cameras; the cameras would be aimed at the faces of customers, so pictures of customers could be obtained. The camera would be the primary means of identifying the customer, and if visual identification failed, the customer could complete the purchase by using a PIN (personal identification number). After the cash register had calculated the total sale, the face recognition system would verify the identity of the customer and the total amount of the sale would be deducted from the customer's bank account. Hence, face-based retailing would provide convenience for retail customers, since they could go shopping simply by showing their faces, and there would be no need to bring debit cards, or other financial media. Wide-reaching applications of face-based retailing are possible, including retail stores, restaurants, movie theaters, car rental companies, hotels, etc.e.g. Swiss European surveillance: facial recognition and vehicle make, model, color and license plate reader.

### IV. SCOPE IN INDIA

1. In order to prevent the frauds of ATM in India, it is recommended to prepare the database of all ATM customers with the banks in India & deployment of high resolution camera and face recognition software at all ATMs. So, whenever user will enter in ATM his photograph will be taken to permit the access after it is being matched with stored photo from the database.

2. Duplicate voter are being reported in India. To prevent this, a database of all voters, of course, of all constituencies, is recommended to be prepared. Then at the time of voting the resolution camera and face recognition equipped of voting site will accept a subject face 100% and generates the recognition for voting if match is found.

3. Passport and visa verification can also be done using face recognition technology as explained above.

4. Driving license verification can also be exercised face recognition technology as mentioned earlier.

5. To identify and verify terrorists at airports, railway stations and malls the face recognition technology will be the best choice in India as compared with other biometric technologies since other technologies cannot be helpful in crowdy places.

6. In defense ministry and all other important places the face technology can be deployed for better security.

7. This technology can also be used effectively in various important examinations such as SSC, HSC, Medical, Engineering, MCA, MBA, B- Pharmacy, Nursing courses etc. The examinee can be identified and verified using Face Recognition Technique.

8. In all government and private offices this system can be deployed for identification, verification and attendance.

9. It can also be deployed in police station to identify and verify the criminals.

10. It can also be deployed vaults and lockers in banks for access control verification and identification of authentic users.

11. Present bar code system could be completely replaced with the face recognition technology as it is a better choice for access & security since the barcode could be stolen by anybody else.

### V. CONCLUSIONS

Face recognition technologies have been associated generally with very costly top secure applications. Today the core technologies have evolved and the cost of equipments is going down dramatically due to the integration and the increasing processing power. Certain applications of face recognition technology are now cost effective, reliable and highly accurate. As a result there are no technological or financial barriers for stepping from the pilot project to widespread deployment.

Though there are some weaknesses of facial recognition system, there is a tremendous scope in India. This system can be effectively used in ATM's ,identifying duplicate voters, passport and visa verification, driving license verification, in defense, competitive and other exams, in governments and private sectors.

Government and NGOs should concentrate and promote applications of facial recognition system in India in various fields by giving economical support and appreciation.

### ACKNOWLEDGMENT

We are thankful to Hon. Ashokrao Chavan (Chief Minister, Maharashtra) India, Society members of Shri. Sharada Bhawan Education Society, Nanded. Also thankful to Shri. Jaiprakash Dandegaonkar (Ex-State Minister, Maharashtra), Society members of Bahiri Smarak Vidyalya Education Society, Wapti for encouraging our work and giving us support.

Also thankful to our family members and our students.

## AUTHORS PROFILE

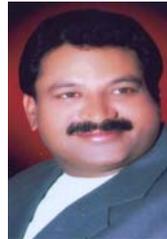

**Dr.S.B.THorat**
M.E. (Computer Science & Engg.)
M.Sc. (ECN), AMIE, LM-ISTE, Ph.D. (Comp.Sc. & Engg.)

He is having 24 years teaching experience. From 2001 he is working as a Director, at ITM. He is Dean of faculty of Computer studies at Swami Ramanand Teerth Marathwada University, Nanded (Maharashtra). Recently he is completed his Ph.D. He attended many national and International conferences. He is having 8 international publications. His interested area are AI, Neural network, Data mining, Fuzzy systems, Image processing.

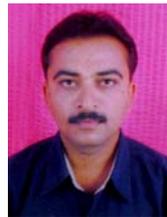

**S.K.Nayak**
M.Sc. (Computer Science), D.B.M, B.Ed.

He completed M.Sc. (Computer Science) from S.R.T.M.U, Nanded. In 2000 he joined as lecturer in Computer Science at Bahirji Smarak Mahavidyalaya, Basmathnagar. From 2002 he is acting as a Head of Computer Science department. He is doing Ph.D. He attended many national and international conferences, workshops and seminars. He is having 7 international publications. His interested areas are ICT, Rural development, Bioinformatics.

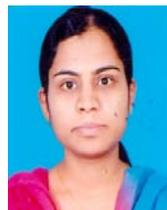

**Miss.Jyoti P Dandale**
B.E. (Computer Science & Engg.)

She has comleted BE from SSGMCE Shegaon. Since 2 years,She has been working as a  lecturer. She has Presented Paper in International conference. She is having 1 international publication. Her interested areas are data and internet security.